\documentclass[letterpaper, 10 pt, conference]{ieeeconf}  

\IEEEoverridecommandlockouts                              

\pdfoutput=1 
\usepackage{graphics} 
\usepackage{epsfig} 
\usepackage{amsmath} 
\usepackage{amssymb}  

\usepackage{xcolor}

\DeclareMathOperator{\sign}{sgn}

\title{\LARGE \bf
Transition control of a tail-sitter UAV using recurrent neural networks
}

\author{Alejandro Flores, and Gerardo Flores
\thanks{Perception and Robotics Laboratory, Centro de Investigaciones en \'Optica A.C., 37150, Le\'on, Guanajuato, M\'exico. Emails: alejandrofl@cio.mx, gflores@cio.mx}%
}

\begin{document}

\maketitle
\thispagestyle{empty}
\pagestyle{empty}

\begin{abstract}               
This paper presents the implementation of a Recurrent Neural Network (RNN) based-controller for the stabilization of the flight transition maneuver (hover-cruise and vice versa) of a tail-sitter UAV. The control strategy is based on attitude and velocity stabilization. For that aim, the RNN is used for the estimation of high nonlinear aerodynamic terms during the transition stage. Then, this estimate is used together with a feedback linearization technique for stabilizing the entire system. Results show convergence of linear velocities and the pitch angle during the transition maneuver. To analyze the performance of our proposed control strategy, we present simulations for the transition from hover to cruise and vice versa. 
\end{abstract}

\section{Introduction}
\begin{figure}[t!]
    \centering 
    \includegraphics[width = \columnwidth]{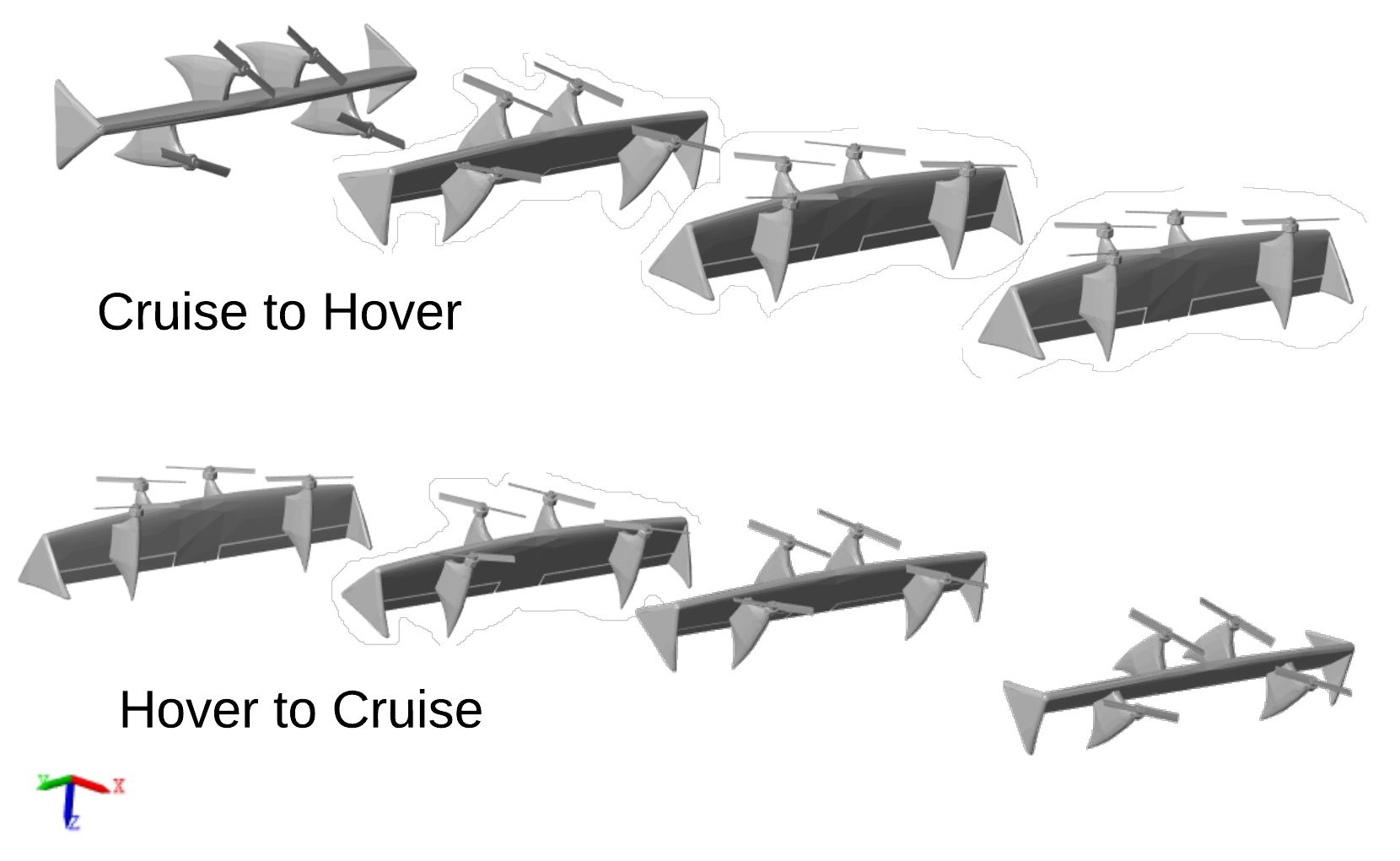}
    \caption{Representation of the two flight modes during the transition maneuver in a tail-sitter drone. For simulation result video see \url{https://youtu.be/frN5Bcow9xw}}
    \label{fig:main_fig}
\end{figure}

The development and application of new Unmanned Aerial Vehicles (UAV's), i.e. drones with hybrid flight capabilities has created a new research area on flight stability and control called, the problem of stabilizing the transition maneuver; which consists in passing from hover flight mode (static flight), to cruise mode (high-speed flight), and vice-versa \cite{7040348}. In the case of tail-sitter aerial vehicles, such transition is achieved by the whole rotation of the UAV body in their $y$-axis, as can be seen in the fig. \ref{fig:main_fig}. Given this and the rise in the use of ANN, several research topics have been carried out with the issue of UAV control systems using ANN.
	
The concept of Artificial Neural Networks (ANN) arose approximately in 1940, and in 1958 the first functional ANN with multiple layers was created. Ever since Neural Networks (NN) have been developed and used in a wide range of applications in robotics, vision, and control systems. Currently, because of the computing capabilities are greater than those of a few decades ago, the implementation of ANN's has increased considerably by the ease in designing and training such networks. 
The application of ANN for control of a wide range of dynamic systems is an actual area of research. In \cite{8790423,6565086,8625222}, authors use Radial Basis Function Neural Network (RBFNN) together with a common sliding mode and PID controller in UAVs, where the NN helps to deal with the nonlinearities, getting good performance tracking in both simulations and real tests. On the other hand, in \cite{8767896,8287305,8407041}, a different RBFNN approach is applied in a quadrotor, where the NN is designed and trained with the main purpose of estimating uncertainties and disturbance moments. Another use for neural networks in dynamical systems is shown in \cite{doi:10.2514/6.2019-1596}, here a neural network was trained to be used as a tool for the estimation of wind velocity based on the information given by a quadcopter, such as trajectory and position. In \cite{7525104}, \cite{4560239,339084,857844} authors worked on feedback linearization based neural network for dynamical systems. The estimation of attitude using ANN is also studied, in \cite{8489118,8168502} authors proposed a long and short term memory neural network (LSTMNN) and a  Modified Elman Recurrent Neural Network (MERNN) respectively used to control attitude and altitude of UAVs.

Concerning flight transition methods, nowadays exist different transition approaches which include from a simple switch between flight modes such as \cite{flores_cuav_jirs} and \cite{gerardo:CDC12}, or continuous strategies \cite{Flores:IFAC14}. In \cite{doi:10.2514/1.C033339} instead of proposing a control for transition, authors design an algorithm to generate the optimal trajectories the UAV should perform for a fast and secure transition flight mode. In \cite{doi:10.2514/1.G003201} it is proposed a unified controller that governs the UAV system in all flight modes, including the transition phase. Also in \cite{doi:10.2514/1.G004697}, it is presented a second-order control law based on proportional derivative errors. Flight mode transition based in Multiple-Model Adaptive Control (MMAC) was applied in some research documents like in \cite{doi:10.2514/1.C034232}. In \cite{doi:10.2514/6.2007-2752} a linear quadratic regulator (LQR) together with a NN, are used to generate a desired command in pitch angle for the transition control of a tail-sitter drone. Finally, in \cite{doi:10.2514/1.29261} authors describe a dynamic inversion and a NN application to perform the transition of a fixed-wing UAV between flight modes. 
 
The objective of this research is to implement an RNN that estimates the nonlinearities presented thanks to abrupt changes in aerodynamics. Once the RNN has estimated such nonlinearities, it is performed a feedback linearization control to perform the transition of the tail-sitter UAV, this by only using information of velocity. To the best of our knowledge, this works is the first that combines RNN with feedback linearization to solve the transition control of tail-sitter vehicles. 

This paper is structured as follows, section \ref{sec:problem_form} describes the system model and a brief introduction to the RNN, section \ref{sec:main_res} presents the principal result obtained by training and implementing the RNN, also a stability analysis of the RNN training and tail-sitter system are described. Next in section \ref{sec:simulations} are exposed and explained the simulations performed to show the control and the system functionality. Finally, conclusion and future work are described in section \ref{sec:conclusion}.
\section{Problem Formulation}
\label{sec:problem_form}
In this section we present the tail-sitter UAV mathematical model, the RNN model, and the problem statement.
\subsection{System Model}
\begin{figure}
	\centering 
	\includegraphics[width = 0.8\columnwidth]{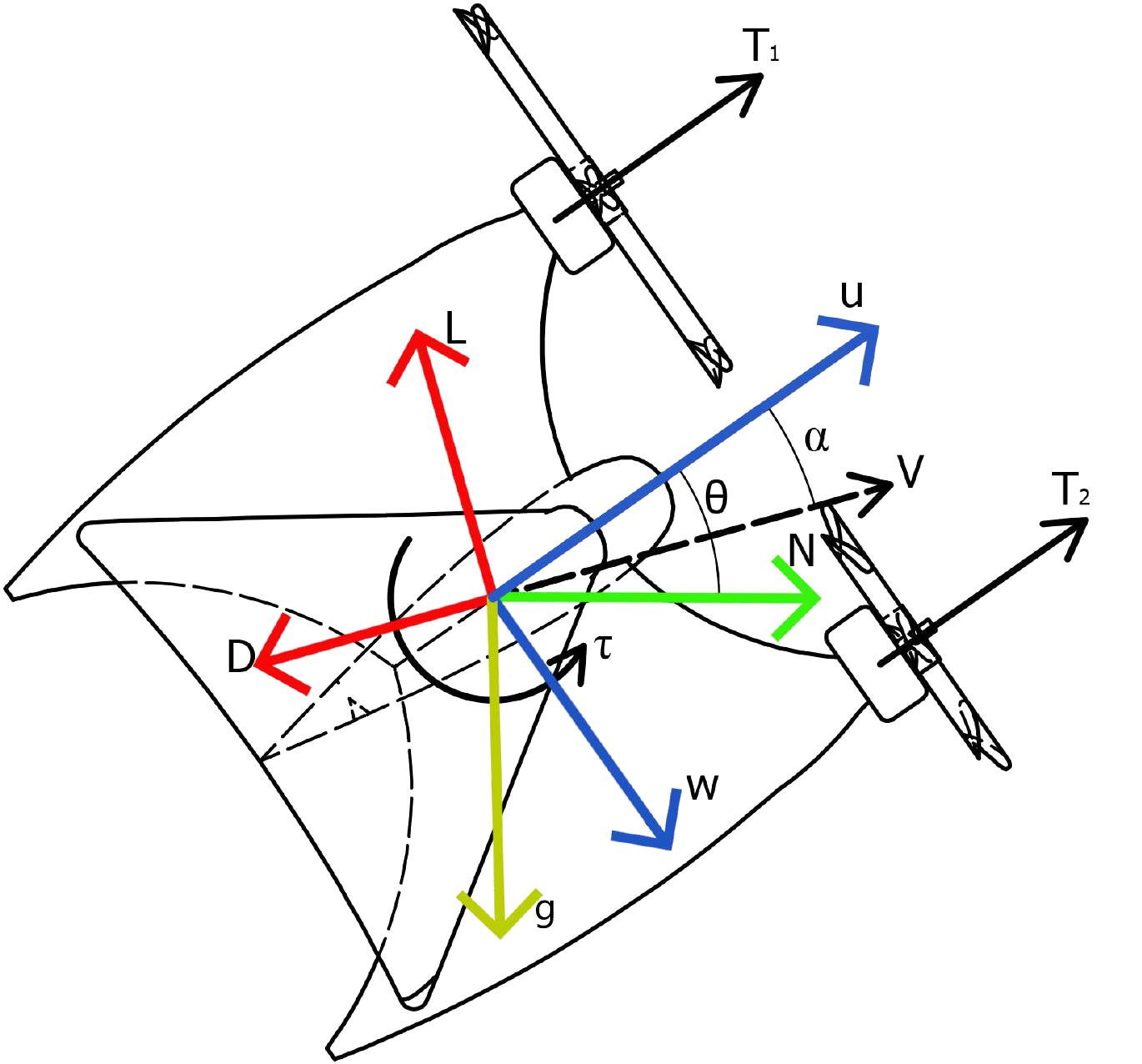}
	\caption{System forces taken into account for mathematical modeling and control design.}
	\label{fig:forces}
\end{figure}

Giving continuity to our work \cite{8619303}, we will use the same system shown in fig. \ref{fig:forces}, where the mathematical model will be defined in the ($x,z$)-body plane as longitudinal dynamics \eqref{eq:sum1} and attitude dynamic \eqref{eq:sum2}
\begin{eqnarray}
\label{eq:sum1}
&\Sigma_{1}&\begin{cases}
        	\dot{u} =\frac{1}{m} \left( T -D \cos\alpha + L\sin\alpha \right) - g\sin\theta - qw  \\
        	\dot{w} = \frac{1}{m}\left( -D \sin\alpha - L\cos\alpha \right) + g\cos\theta + qu 
	\end{cases}\\
\label{eq:sum2}
    &\Sigma_{2}& \begin{cases}
        	\dot{\theta} = q\\
        	\dot{q} = \frac{1}{J} \tau
	\end{cases}
\end{eqnarray}
where $u$ and $w$ are the vertical and horizontal body speeds; $D$ and $L$ are the drag and lift aerodynamic forces, respectively; $\theta$ and $q$ are the pitch angle and rate of the UAV, respectively; $\alpha$ represents the UAV's angle of attack (AoA); $m$ and $J$ represent the UAV mass and its inertia in the $y$-axis, respectively; and $\tau$, $T$ are the pitching moment and thrust, both considered as control inputs.

Since $\dot{w}$ in \eqref{eq:sum1} is not directly controlled, but it depends on states $u$, $\theta$ and $q$ it is important to define appropriate desired values to keep $w$ stable. To achieve this, we propose to use $\theta$ as a virtual controller such that
\begin{equation}
    \epsilon = \cos(\theta),
    \label{eq:virtual_control}
\end{equation}
now defining $h_{1}(u,w,q) = \frac{1}{m}( -D \cos\alpha + L\sin\alpha ) - qw$ and $h_{2}(u,w,q) = \frac{1}{m}( -D \sin\alpha - L\cos\alpha ) + qu$, then  \eqref{eq:sum1} leads to
\begin{eqnarray}
\label{eq:u_f1}
    \dot{u} &=& h_{1}(u,w,q) - g\sqrt{1 - \epsilon^2} + \frac{T}{m}\\
    \label{eq:w_f2}
    \dot{w} &=& h_{2}(u,w,q) + g\epsilon
\end{eqnarray}
then we search for controllers
\begin{eqnarray}
\label{eq:control1}
    T &=& -m \left( \hat{h}_{1}(\cdot) + \upsilon_u \right) + g\sqrt{1 - \epsilon^2} \\
    \label{eq:control2}
    \epsilon &=& - \frac{1}{g} \left( \hat{h}_{2}(\cdot) + \upsilon_w \right)
\end{eqnarray}
where $(\hat{h}_{1}(\cdot), \hat{h}_{2}(\cdot))$ are estimates of the real functions $( h_{1}(u,w,q), h_{2}(u,w,q) )$. Such estimates come from the output of the RNN. If the estimates $(\hat{h}_{1}, \hat{h}_{2})$ converge to the real values $({h}_{1}, {h}_{2})$, the system \eqref{eq:u_f1}-\eqref{eq:w_f2} with previous controllers would have the form
\begin{eqnarray}
    \label{eq:u_linear}
    \dot{u} &=& \upsilon_u\\
    \label{eq:w_linear}
    \dot{w} &=& \upsilon_w
\end{eqnarray}
where $\upsilon_u$ and $\upsilon_w$ could be designed as standard proportional controllers of the form $\upsilon_u = -k_1(u - u_d)$ and $\upsilon_w = -k_2(w - w_d)$ with $k_1$ and $k_2$ being positive constants.  As we can see in \eqref{eq:u_f1}, the control input $\epsilon$ should be in the rank of $\pm 1$. The way we can obtain estimates of $h_1, h_2$ is through the implementation of RNN explained next.

\subsection{Recurrent Neural Network}
An RNN is a type of neural network with feedback loops in their neurons, making its output a function of both the actual state and its inputs. Its structure permits estimating nonlinear function in dynamic systems. This type of ANN is constructed in layers consisting of: weight matrices, internal and external inputs, transfer functions, and outputs. A mathematical model of an RNN can be expressed as follows
\begin{equation}
    \dot{\textbf{x}} = -\textbf{Cx} + \textbf{W}_{x}\textbf{f}(\textbf{x}) + \textbf{W}_{p}\textbf{f}(\textbf{p})
    \label{eq:dynamic_ANN}
\end{equation}
where $\textbf{x}$ is the RNN output vector; $\textbf{f}(\cdot)$ is a bounded and non-decreasing activation function of the neuron, in this case hyperbolic tangent function; $\textbf{C}$ and $\textbf{W}_x$ are the connection weight matrices with $\textbf{C} = diag\{ c_1,c_2, ..., c_n\}$ and $c_i > 0$; $\textbf{W}_p$ is the external input weight matrix; and $\textbf{p}$ is the external ANN input vector.

\subsection{Problem Statement} 
The objective of this work is to perform a stable transition of the tail-sitter UAV by controlling the aircraft velocity and the pitch angle. For that, we follow three steps:
 \begin{enumerate}
     \item Consider the pitch angle $\theta$ as a virtual control as indicated in \eqref{eq:virtual_control}. With such a virtual control the $w$ dynamics are controlled. The output of this virtual control is $\theta_d$.
     \item Train two RNN to estimate the nonlinear components $h_1$ and $h_2$ in \eqref{eq:u_f1} and \eqref{eq:w_f2} respectively. Once these functions are estimated, we apply a feedback linearization controller.
     \item Once we have $\theta_d$ from step 1, this serves as a reference for pitch controller $\tau$ in the pitch dynamics $\Sigma_2$.
 \end{enumerate}
With the previous steps, we can achieve a stable flight transition of a tail-sitter UAV. This controller approach will be tested with different initial conditions to ensure the functionality in the system.
\section{Main Result}
\label{sec:main_res}
This section describes the training procedure of the RNN, as well as the control algorithm together with the definition of desired trajectories. Also, the stability analysis is driven showing that the system converges to the desired values.

\subsection{RNN Training}
\begin{figure}
    \centering 
    \includegraphics[width = \columnwidth]{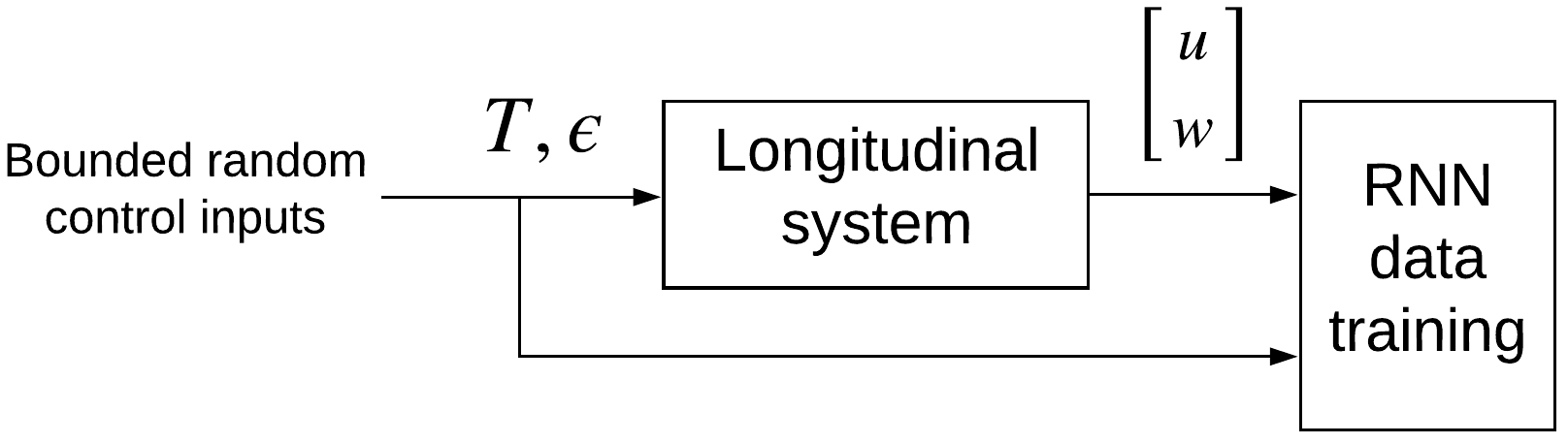}
    \caption{Block diagram used for getting the training data for the RNN. The longitudinal system dynamics are equation \eqref{eq:u_f1} and \eqref{eq:w_f2}}
    \label{fig:LDRN_block_training}
\end{figure}

\begin{figure}
    \centering
    \includegraphics[width = 0.9\columnwidth]{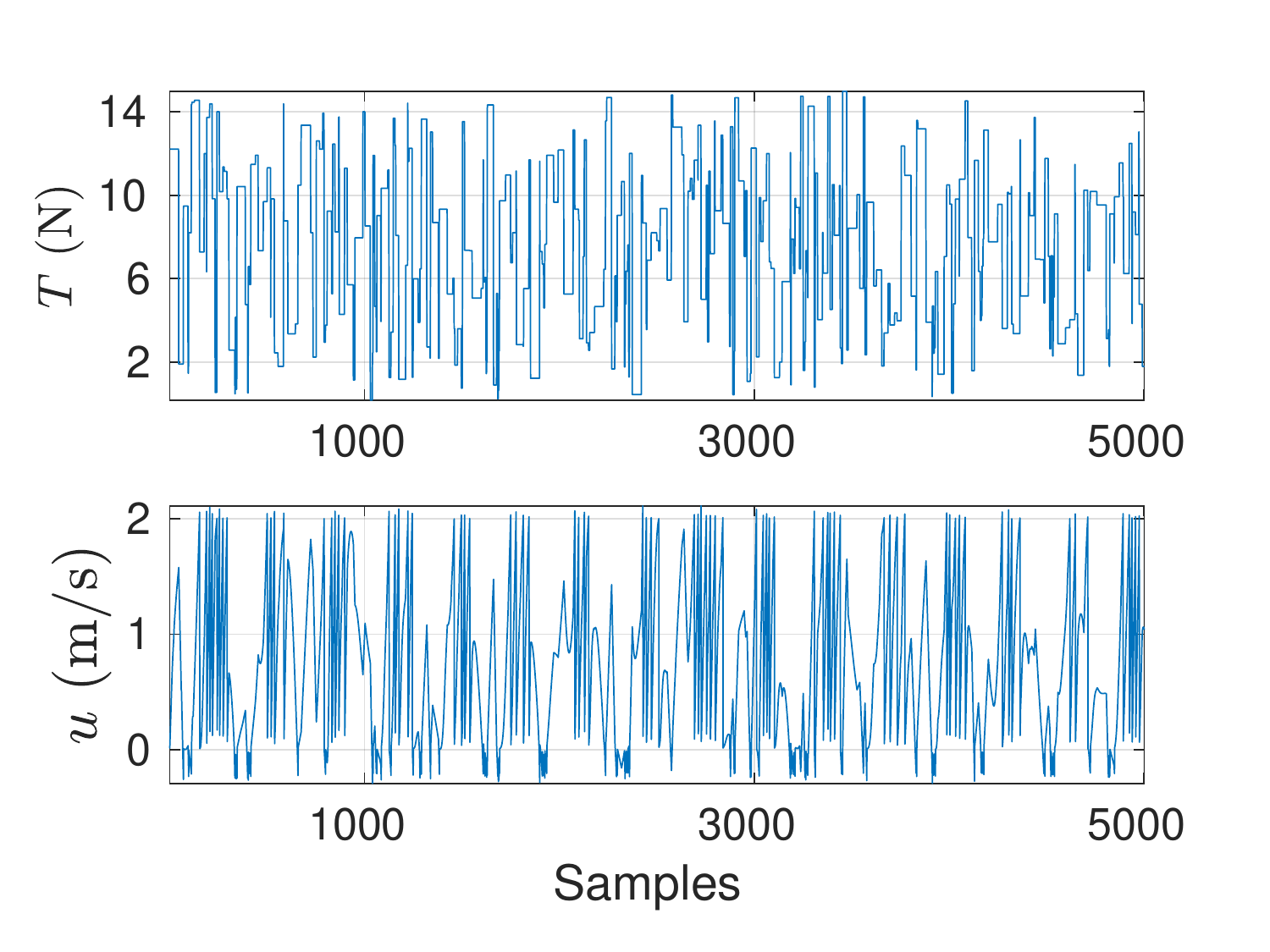}
    \caption{Control input $T$ applied randomly to system \eqref{eq:u_f1}, and velocity $u$ obtained from system during a simulation. This data is used for the first RNN training.}
    \label{fig:data_thrust}
\end{figure}

\begin{figure}
    \centering
    \includegraphics[width = 0.9\columnwidth]{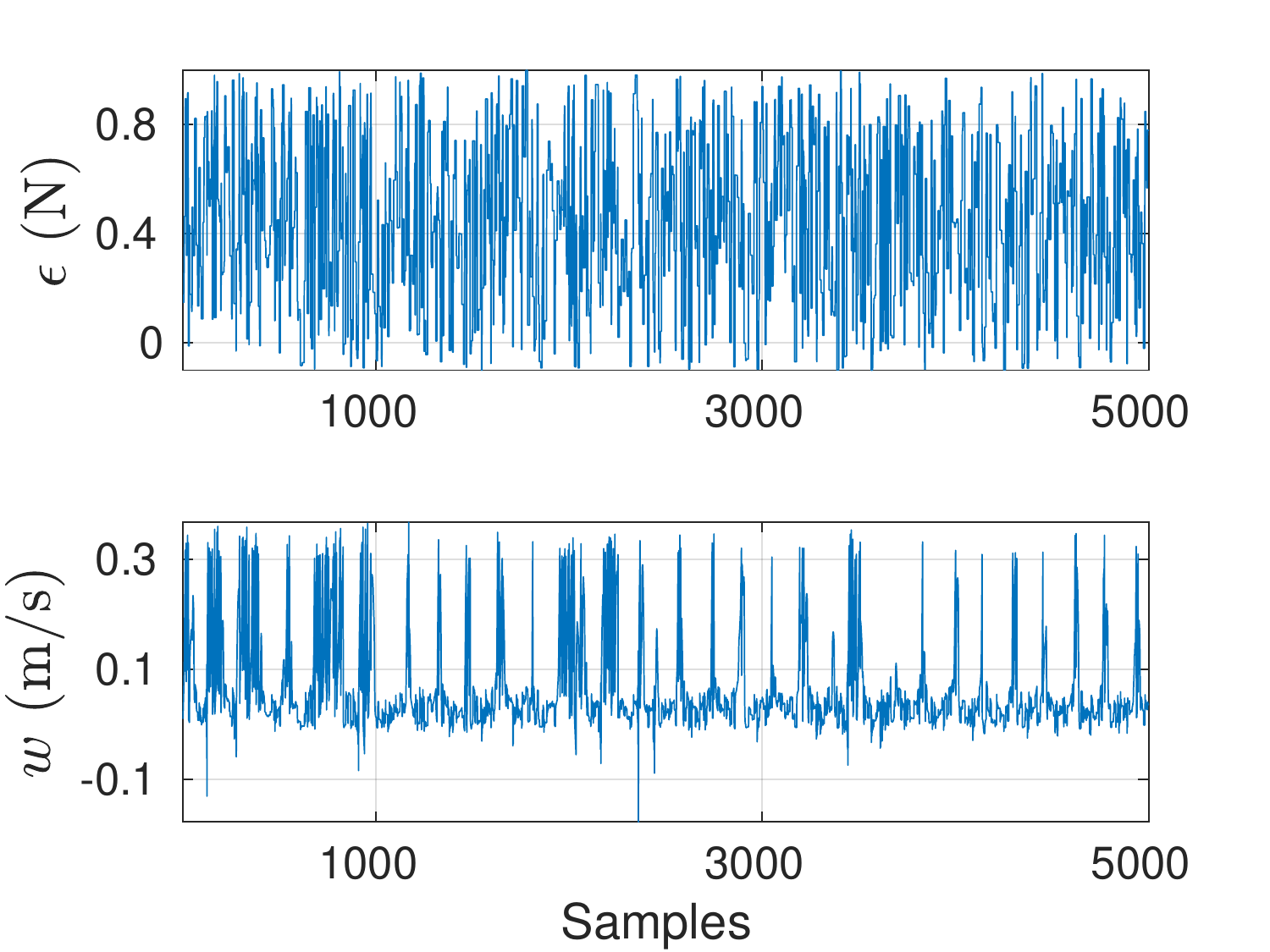}
    \caption{Control input $\epsilon$ applied randomly to system \eqref{eq:w_f2}, and velocity $w$ obtained from system during a simulation. This data is used for the second RNN training.}
    \label{fig:data_epsilon}
\end{figure}

To train the RNN, it is necessary to define the mathematical model of the network to approximate its output. The input of the RNN is given by random control inputs $( \epsilon , T)$. A scheme of the RNN is depicted in fig. \ref{fig:LDRN_block_training}.
The training data-set comes from a simulation of the system applying random bounded input values to the system, and observe how the system responds. The data obtained while performing simulation can be seen in fig. \ref{fig:data_thrust} and fig. \ref{fig:data_epsilon} where 5000 samples were collected for each NN, in these graphs we can see that random control inputs $T$ and $\epsilon$ (upper graphs) applied to \eqref{eq:u_f1} and \eqref{eq:w_f2} respectively produce the behavior on $u$ and $w$ shown in the lower graphs. With this data-set we can start training the NN's.

\begin{figure}
    \centering
    \includegraphics[width =0.9\columnwidth]{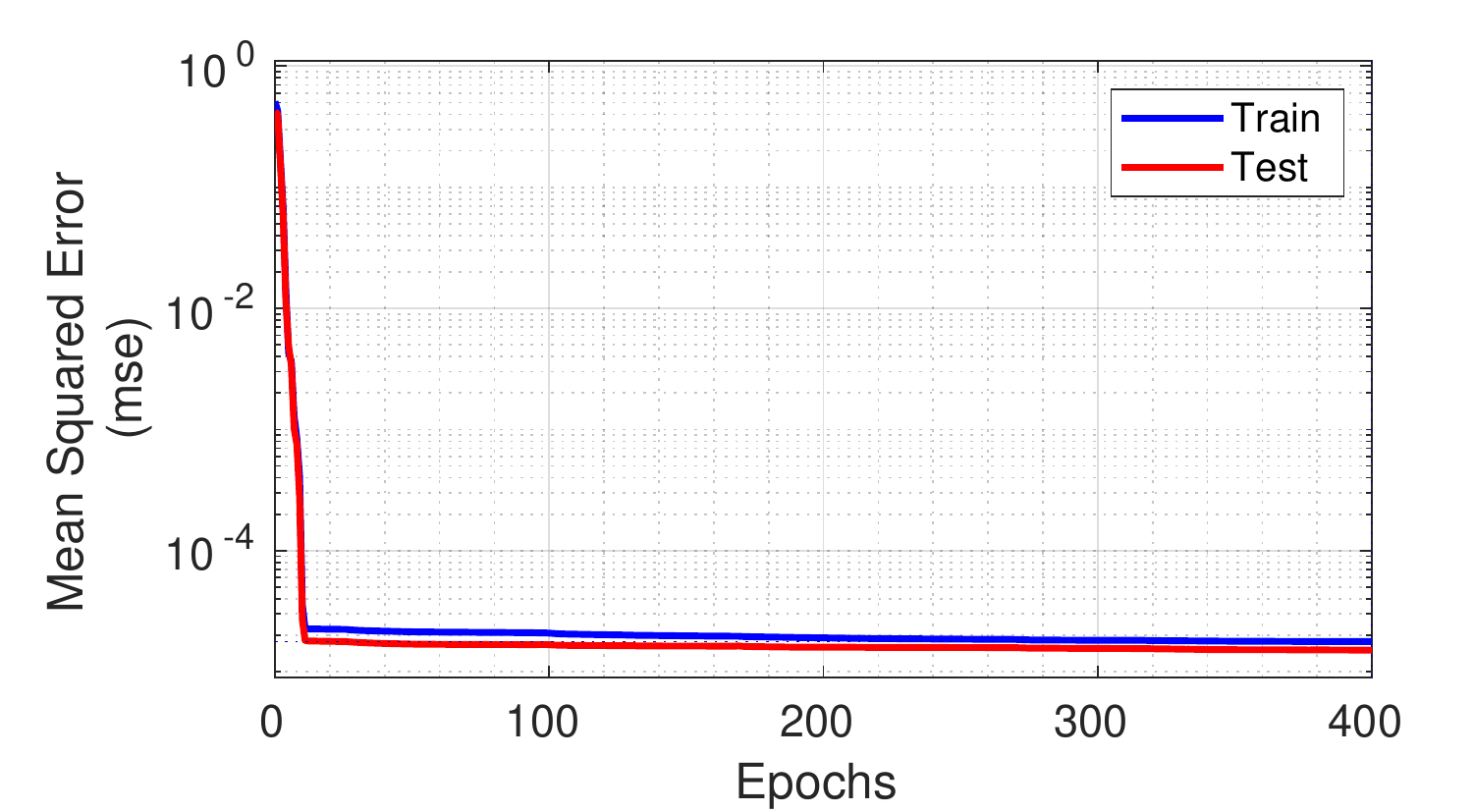}
    \caption{Performance obtained from the first RNN training ($h_1$ estimation). This graphs shows the performance in terms of error during the training.}
    \label{fig:train_performance}
\end{figure}

\begin{figure}
    \centering
    \includegraphics[width = \columnwidth]{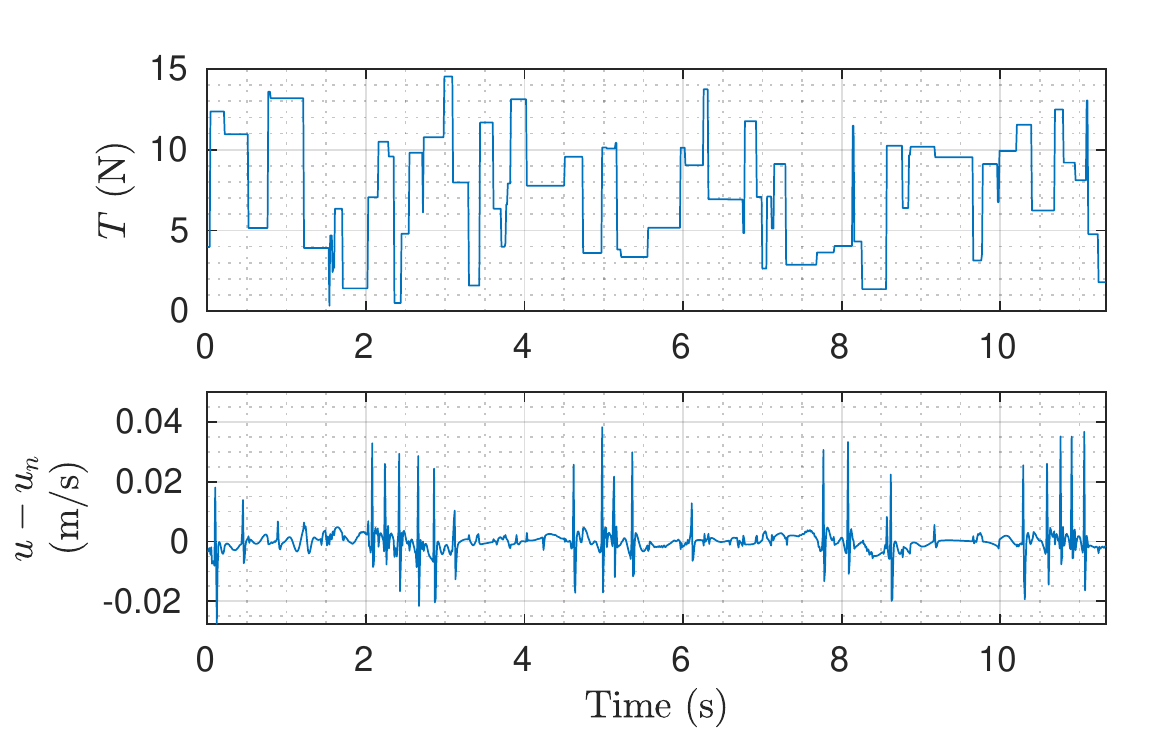}
    \caption{Error obtained from the trained RNN (lower graph) during a simulation of the system applying random input $T$ (upper graph). This error is defined as the real system response and the RNN estimation output.}
    \label{fig:train_test}
\end{figure}

\begin{figure}
    \centering
    \includegraphics[width = \columnwidth]{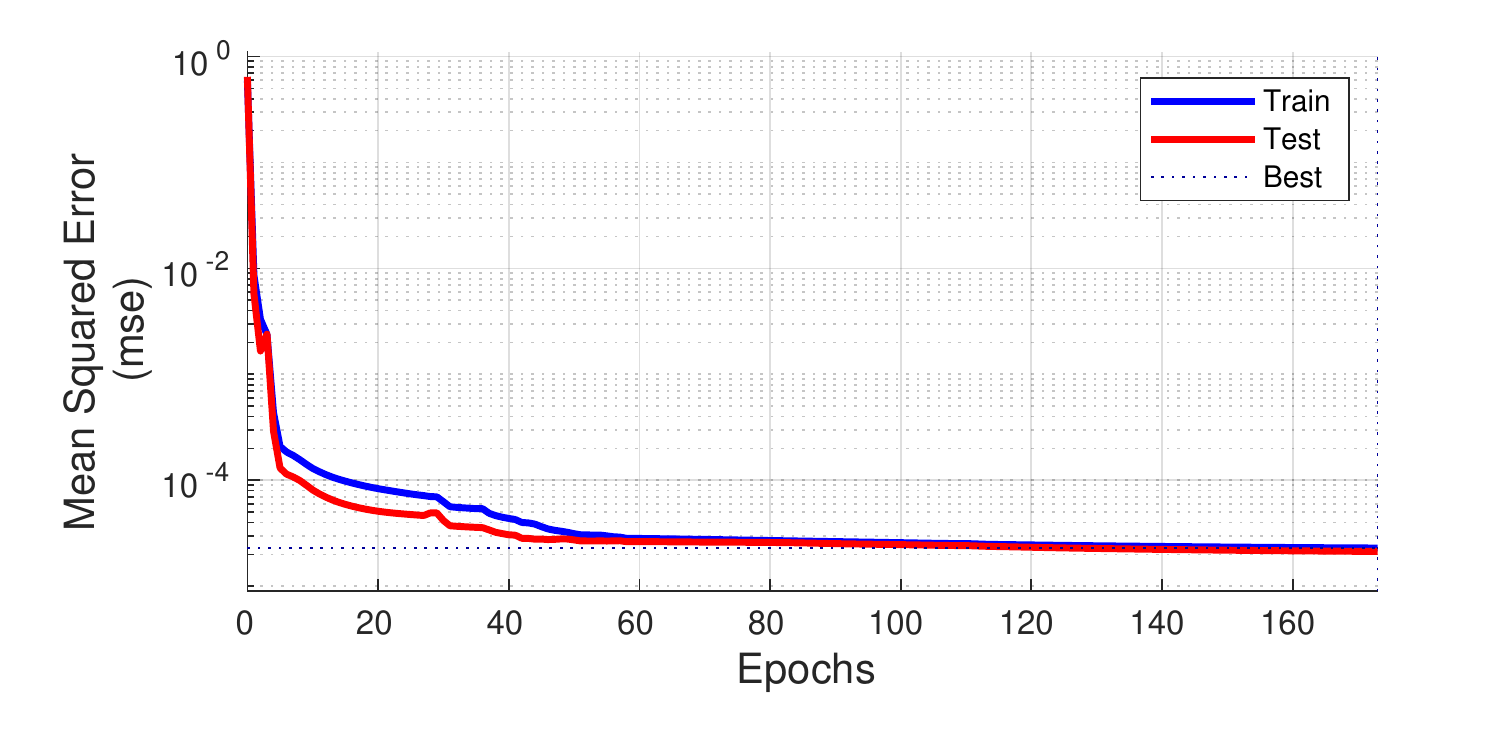}
    \caption{Performance obtained from the first RNN training ($h_2$ estimation). This graphs shows the performance in terms of error during the training.}
    \label{fig:train_performance_epsilon}
\end{figure}

\begin{figure}
    \centering
    \includegraphics[width = \columnwidth]{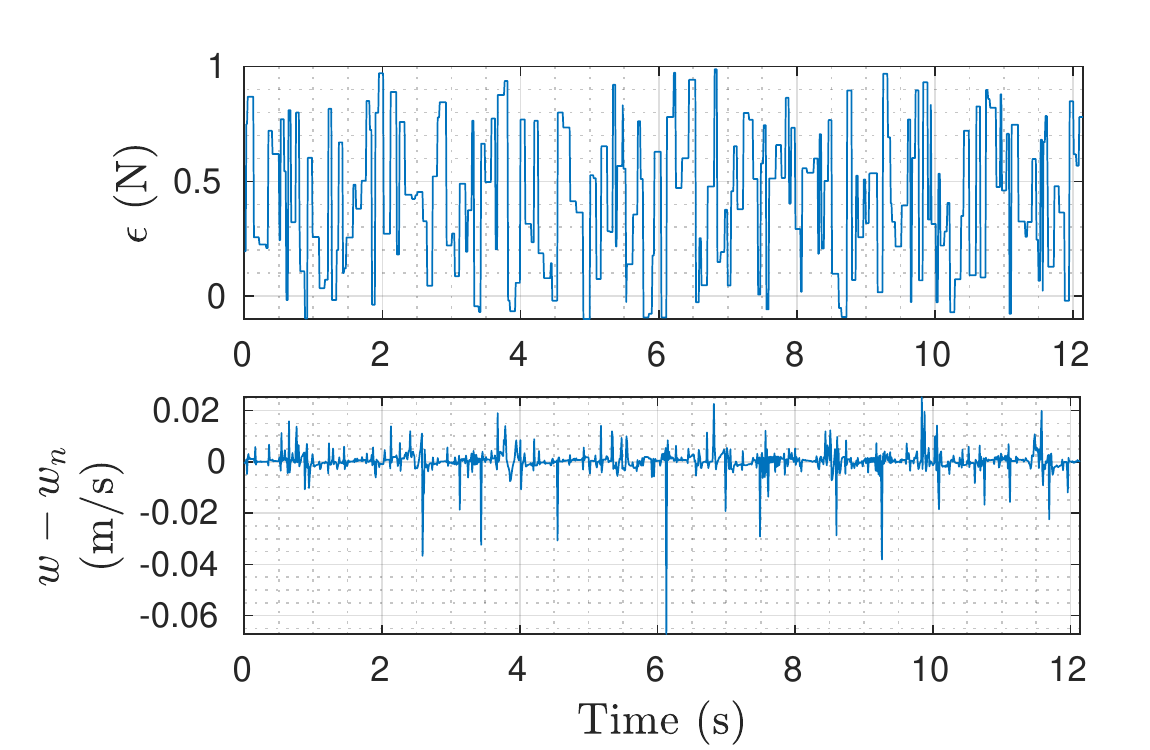}
    \caption{Error obtained from the trained RNN (lower graph) during a simulation of the system applying random input $\epsilon$ (upper graph). This error is defined as the real system response and the RNN estimation output.}
    \label{fig:train_test_epsilon}
\end{figure}

After training the RNN's it is important to analyze the performance obtained in this process. For doing this we define a mean square error (MSE) between the real system output ($u$) and the output estimated by the NN ($u_n$) as
\begin{equation*}
    MSE = \frac{1}{m} \sum_{}^{m}(u - u_n)^2
\end{equation*}
where m is the number of samples, in this case 5000. The results obtained in a training of 400 epochs are shown in fig. \ref{fig:train_performance} where it can be seen that the mean squared error is relatively low at the end of the training meaning that the approximation of $h_{1}$ is acceptable. To show this error, a simple test was performed to measure the error obtained between the NN and the real system behaviors. Fig. \ref{fig:train_test} depicts the results obtained, this graph shows that, compared with the drastically control input changes, the error obtained is low. Same procedure were perform for the second RNN to estimate $h_2$ and the results are shown in fig. \ref{fig:train_performance_epsilon} and fig. \ref{fig:train_test_epsilon} respectively. 
\subsection{Design of the system desired values}
Since subsystem \eqref{eq:sum1} has slower dynamics than the attitude subsystem \eqref{eq:sum2} \cite{gerardo:ACC13}, we will focus designing the desired velocity values $(u_{d}, w_{d})$ to ensure that the UAV will keep a secure altitude during the flight mode transition. In this process, the state $\theta$ will serve as virtual control for stabilizing the dynamics of $w$ to its desired value $w_d$. Then, for the transition maneuver from hover to cruise, $u$ must increase in such a way that the lift force is enough to compensate the aircraft's mass and gravity. While $w$ must be designed w.r.t. $u$ to have the aircraft AoA at the optimal value considering the lift-drag ratio $(L/D)$. $L$ and $D$ are defined as $D = KC_DV^2$ and $L = KC_LV^2$; where $K$ is a positive constant; $C_L$ and $C_D$ are the lift and drag coefficients, respectively; and $V^2 = \sqrt{u^2 + w^2}$ is the aircraft airspeed. In this case $\alpha_d$ is designed is such a way that its value reaches the optimal AoA according to the transition flight mode. Then, desired values $(u_{d}, w_{d},\alpha_d)$ for transition from hover to cruise, are computed as follows
\begin{eqnarray}
\label{eq:ud}
&u^{hc}_{d}=&\begin{cases}
        	\frac{\arctan \big(a_{u}(\frac{t}{5}-L_{u})\big) }{a_{u}}+L_{u}  &$if$ \hspace{8pt} \frac{t}{5} > L_{u}\\
            \frac{t}{5}   &$if$ \hspace{8pt} \frac{t}{5} \leq L_{u}
	\end{cases}\\
    \label{eq:al}
&\alpha^{hc}_{d}=&\begin{cases}
        	\frac{\arctan \big(a_{\alpha}(t-L_{\alpha})\big) }{a_{\alpha}}+L_{\alpha}  & $if$ \hspace{8pt} t > L_{\alpha}\\
            t  & $if$ \hspace{8pt} t\leq L_{\alpha}
	\end{cases}\\
    \hspace{8pt}
\label{eq:wd}
    &w^{hc}_{d}=& u_{d}\tan\alpha_{d}
\end{eqnarray}
where $a_{u} = \frac{\pi}{2(M_{u}-L_{u})}$, $a_{\alpha} = \frac{\pi}{2(M_{\alpha}-L_{\alpha})}$; $(M_u, L_u)$, $(M_{\alpha}, L_{\alpha})$ are positive constants which $L_u \leq M_u$, $L_{\alpha} \leq M_{\alpha}$. In the same way, when the transition occurs from cruise to hover, both velocities $(u,w)$ must be reduced following the next desired values
\begin{eqnarray}
    u^{ch}_d &=& 1 - u^{hc}\\
    \alpha^{ch}_d &=& 6 - \alpha^{hc}\\
    w^{ch}_d &=& u^{ch}_{d}\tan\alpha^{ch}_{d}.
\end{eqnarray}
With these desired values and applying the virtual control \eqref{eq:virtual_control}, we could obtain also the desired values for $\theta$ in both transitions. Please refer to \cite{8619303} for more details of this process.
\subsection{RNN stability analysis}
To determine if our RNN \eqref{eq:dynamic_ANN} well estimates the nonlinear function, we will determine if the NN weights ($\textbf{W}_x$,$\textbf{W}_p$) converges to the optimal values. If so, we can ensure that the closed-loop system converges to its desired values. For this, we first define an RNN estimation dynamic as follows
\begin{equation}
    \dot{\hat{x}}_i = -\hat{x}_i + \hat{\textbf{w}}_{x_j} \textbf{f}(\hat{\textbf{x}}) + \hat{\textbf{w}}_{p_j} \textbf{f}(\textbf{p})
\end{equation}
where $x_i$ is the $i$-th neuron of the RNN; $\textbf{w}_{x_j}$ the $j$-th row of $\textbf{W}_x$; and $\textbf{w}_{p_j}$ the $j$-th row of $\textbf{W}_p$. And defining the NN output error as $\Tilde{x} = x - \hat{x}$ and the weights error as $\Tilde{\textbf{w}}_{\{x,p\}} = \textbf{w}_{\{x,p\}} - \hat{\textbf{w}}_{\{x,p\}}$ we have
\begin{equation}
    \dot{\Tilde{x}} = -\Tilde{x} + \textbf{w}_x \textbf{f}(\textbf{x}) - \hat{\textbf{w}}_x \textbf{f}(\hat{\textbf{x}}) + \textbf{w}_p \textbf{f}(\textbf{p}) - \hat{\textbf{w}}_p \textbf{f}(\textbf{p})
    \label{eq:error_dot_rnn}
\end{equation}
Now, as a candidate Lyapunov function we have
\begin{equation*}
    V(\Tilde{x},\Tilde{\textbf{w}}_x,\Tilde{\textbf{w}}_p) = \frac{1}{2}\Tilde{x}^{2} + \frac{1}{2}\Tilde{\textbf{w}}_x\Tilde{\textbf{w}}_x^T + \frac{1}{2}\Tilde{\textbf{w}}_p\Tilde{\textbf{w}}_p^T
\end{equation*}
and its derivative
\begin{equation}
    \dot{V}(\Tilde{x},\Tilde{\textbf{w}}_x,\Tilde{\textbf{w}}_p) = \Tilde{x}\dot{\Tilde{x}} + \Tilde{\textbf{w}}_x\dot{\Tilde{\textbf{w}}}_x^T + \Tilde{\textbf{w}}_p\dot{\Tilde{\textbf{w}}}_p^T,
    \label{eq:d_lyapunov}
\end{equation}
finally defining the weight changes as 
\begin{eqnarray}
    \dot{\hat{\textbf{w}}}_x^T &=& \Tilde{x} \textbf{f}(\hat{\textbf{x}})
    \label{eq:W_changes1}\\
    \dot{\hat{\textbf{w}}}_p^T &=& \Tilde{x} \textbf{f}(\textbf{p}).
    \label{eq:W_changes2}
\end{eqnarray}
Substituting \eqref{eq:error_dot_rnn}, \eqref{eq:W_changes1} and \eqref{eq:W_changes2} in \eqref{eq:d_lyapunov} we obtain
\begin{equation}
    \dot{V}(\Tilde{x}_i,\Tilde{\textbf{w}}_x,\Tilde{\textbf{w}}_p)= -\Tilde{x}^2 - \Tilde{x}\textbf{w}_x (\textbf{f}(\hat{\textbf{x}}) - \textbf{f}(\textbf{x}))
\end{equation}
since $\textbf{f}(\cdot)$ is a non decreasing function, then $\sign{(\Tilde{x})} = -\sign{(\textbf{f}(\hat{\textbf{x}}) - \textbf{f}(\textbf{x}))}$, and ensuring that $\textbf{w}_x$ are strictly positive, we demonstrate that system \eqref{eq:error_dot_rnn} converges to zero asymptotically.  

Now, having concluded that the weights converges to the optimal values for the NN, we can ensure that the RNN well estimates the nonlinear parts $h_1$ and $h_2$ of \eqref{eq:u_f1} and \eqref{eq:w_f2}. Now, we can proceed to analyze the closed-loop system dynamics.

\subsection{Closed-loop system behavior}
Once the RNN was trained in such a way that the performance of the NN is optimal, i.e., it estimates the nonlinear functions $h_1$ and $h_2$ correctly, we can define errors $e_u = (u - u_d)$ and $e_w = (w - w_d)$. From \eqref{eq:u_linear} and \eqref{eq:w_linear}, it is easily seen that the feedback control law  
\begin{eqnarray}
\label{eq:control3}
     \upsilon_u &=& - k_1 e_u + \dot{u}_d  \\
    \label{eq:control4}
    \upsilon_w &=& - k_1 e_w + \dot{w}_d
\end{eqnarray}
substituted in \eqref{eq:control1} and \eqref{eq:control2}, yields the convergence of the error dynamics. From this, it is clear that such error dynamics is asymptotically stable if $k_1$ and $k_2$ are strictly positive constants, i.e., $u \rightarrow u_d$ and $w \rightarrow w_d$ as $t \rightarrow \infty$. 

Next, we can analyze the subsystem \eqref{eq:sum2} redefining the virtual control $\epsilon$ as the desired values for $\theta$ as $\theta_d = \arccos{\epsilon}$. Then rewriting \eqref{eq:sum2} in terms of error and defining the control law $\tau = -k_3e_\theta - k_4e_q$ we have
\begin{equation}
   \dot{\textbf{e}} = \textbf{Ae}
\end{equation}
where $\textbf{e} = [e_{\theta} \hspace{5pt} e_q]^T$ with $e_{\theta} = \theta - \theta_d$ and $e_q = q - q_d$ and $\textbf{A}$ is the matrix
\begin{equation}
   \textbf{A} = 
   \begin{bmatrix}
        0 & 1 \\
        -k_3 & -k_4
    \end{bmatrix},
\end{equation}
which is Hurwitz if $k_3$ and $k_4$ are strictly positive and leads the subsystem \eqref{eq:sum2} to be GES meaning that $\theta \rightarrow \theta_d$, and $q \rightarrow q_d$, this is, the attitude system converges exponentially to the desired values $\theta_d$ and $q_d$ for any initial conditions ($\theta(0)$,$q(0)$).
\section{Simulations}
\label{sec:simulations}
\begin{figure}
    \centering
    \includegraphics[width = \columnwidth]{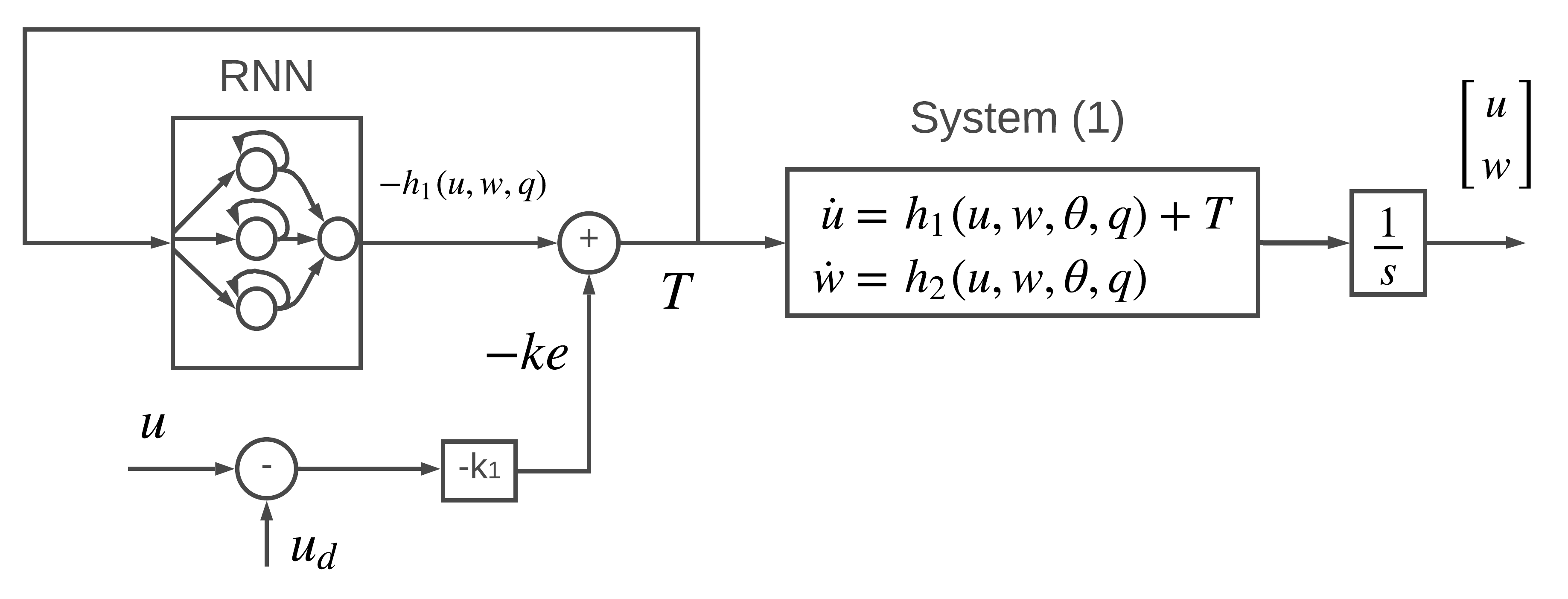}
    \caption{Structure of the longitudinal dynamics when applying the RNN for the feedback linearization and the control law $\sigma = -k_3(u - u_d)$ to achieve desired values.It is important to remember that the RNN estimates the non linear part $h_1$ of $\dot{u}$ dynamics using only the total control input applied to the system}
    \label{fig:RBFNN_block_sim}
\end{figure}

Once the neural network is trained, it is possible to go to the testing stage, in this case, different simulations of the flight mode (hover-cruise) and (cruise-hover) transitions are performed, fig. \ref{fig:RBFNN_block_sim} shows the block diagram used to simulate \eqref{eq:sum1} using the RNN and the control input.  In the next subsections, the results obtained during the simulation of the transitions between the flight modes are explained. In this case we will assume, for simplicity, that the UAV mass and tensor in the $y$-axis are $m = 1$ Kg and $J_{2,2} = 1$ Kg $\cdot$ m$^2$, the lift and drag coefficients were taken from previous work of a symmetric airfoil. It is important to mention that the desired values of $u$ and $w$ were obtained by \eqref{eq:ud} and \eqref{eq:wd} while $\theta_d$ were obtained by the result of the virtual control $\epsilon$ by simulating systems \eqref{eq:u_f1} and \eqref{eq:w_f2} in both flight mode transitions.
\subsection{Hover-Cruise transition}
\begin{figure}
    \centering
    \includegraphics[width = \columnwidth]{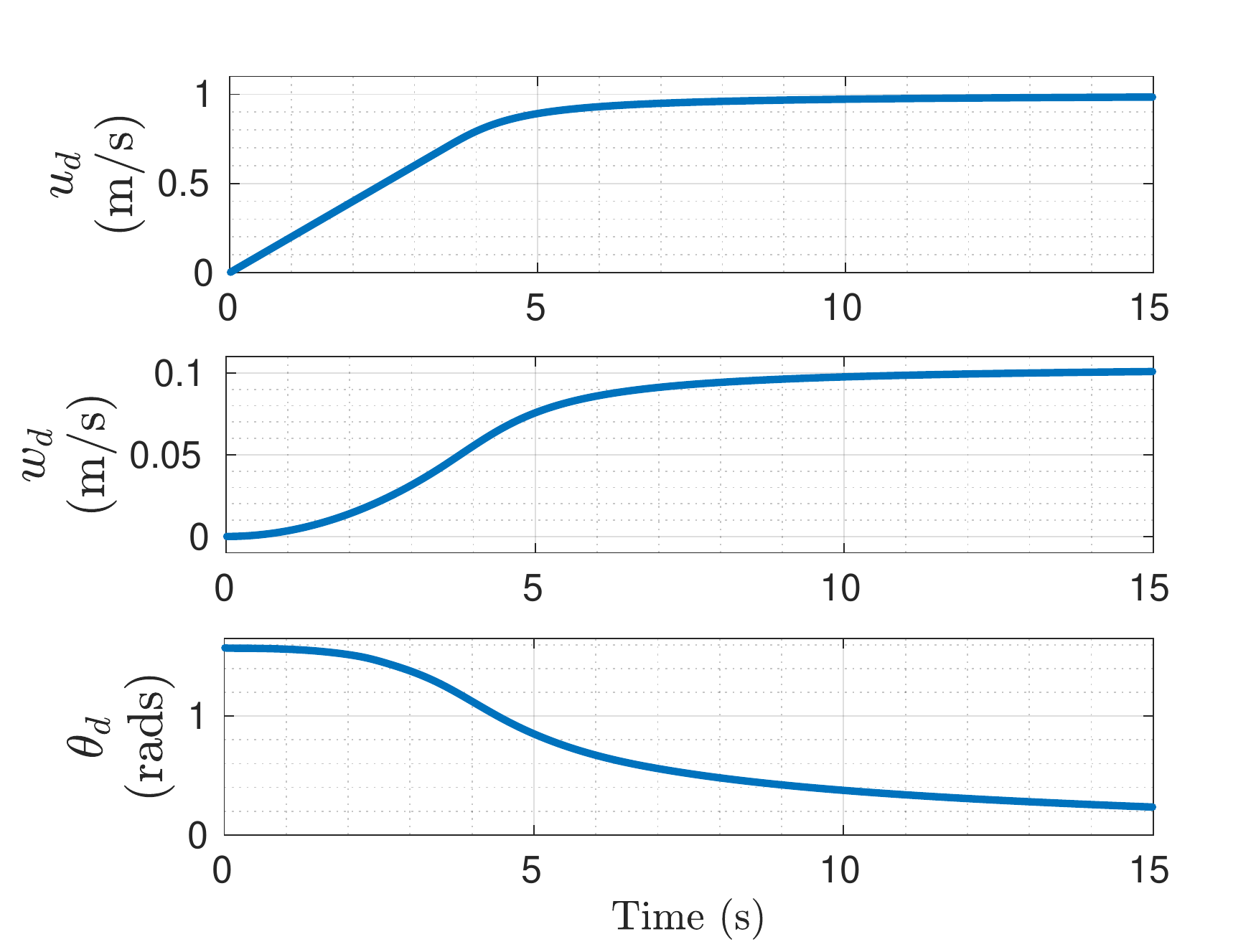}
    \caption{Desired velocities for $u$, $w$ and desired pitch angle $\theta$ during the transition from hover to cruise. As it can be seen, the initial values are approximately zero since the UAV keeps a static flight mode and for transition to cruise, the speed should increase to produce aerodynamic force. This desired values were obtained from equations \eqref{eq:ud} and \eqref{eq:wd}.}
    \label{fig:desired_vel_hc}
\end{figure}

Several simulations were performed for the hover-cruise flight mode transition to ensure the RNN works well, one simulation result is presented next. It is important to determine the desired velocities and pitch angle according to the initial and final flight mode to analyze the convergence of the system and the error presented during simulations. Fig. \ref{fig:desired_vel_hc} shows the desired velocities the UAV must track to keep a good transition and the desired pitch angle.

\begin{figure}
    \centering
    \includegraphics[width = \columnwidth]{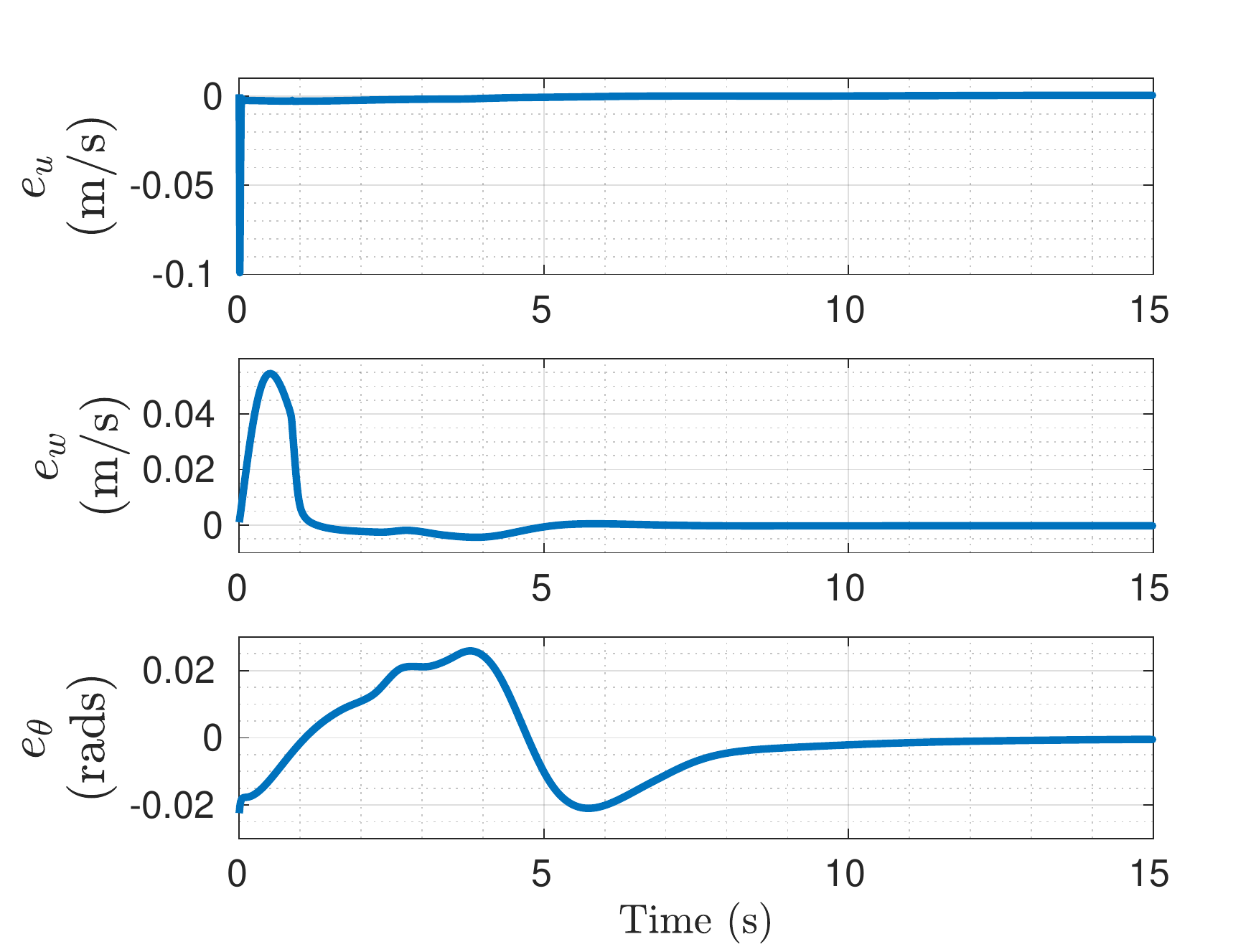}
    \caption{Error velocities in $u$, $w$ and pitch angle error evolution obtained during the simulation for the hover-cruise transition using the trained RNN}
    \label{fig:u_w_hc}
\end{figure}

\begin{figure}
    \centering 
    \includegraphics[width = 0.9\columnwidth]{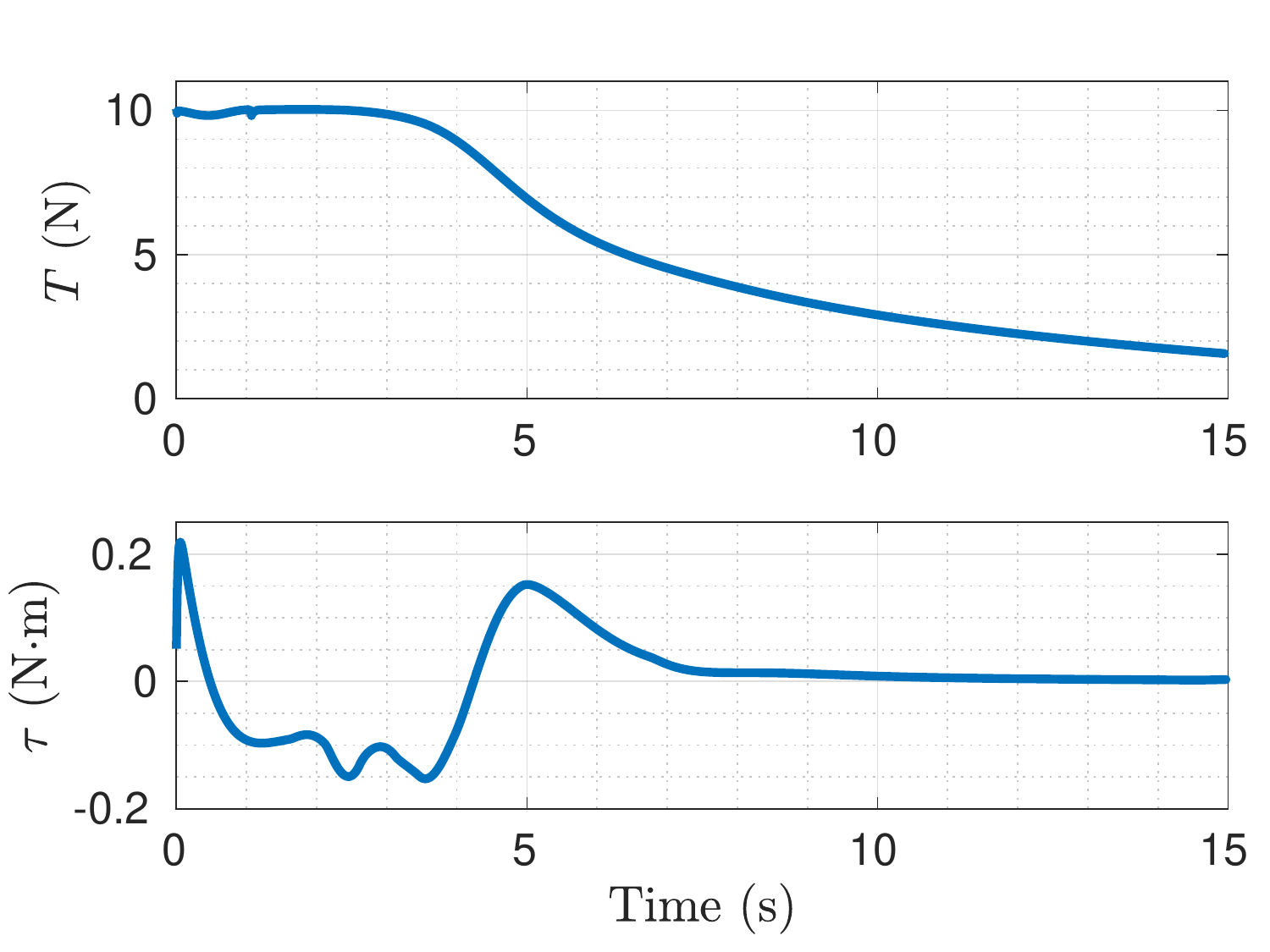}
    \caption{Control input $T$ applied to the system \eqref{eq:sum1} conformed by the computation of the ANN and the proportional controller for the transition from hover to cruise. And control input $\tau$ applied to the subsystem \eqref{eq:sum2} for controlling the pitch angle}
    \label{fig:thrust_hc}
\end{figure}

As the initial flight mode is hover, the initial conditions of the velocities were $u(0)=0.01$ m/s and $w(0)=0.001$ m/s and $\theta(0)=1.6$ rads. The results obtained in the implementation of the NN are shown in fig. \ref{fig:u_w_hc} which depicts the error obtained between the actual and the desired values, also the pitch angle error during simulation is shown. In this simulation the flight mode transition is completed in about 13 seconds, fig. \ref{fig:thrust_hc} shows the control inputs $T$ and $\tau$ applied to the longitudinal system  \eqref{eq:sum1} and the attitude system \eqref{eq:sum2}.
\subsection{Cruise-Hover transition}
\begin{figure}
    \centering
    \includegraphics[width = \columnwidth]{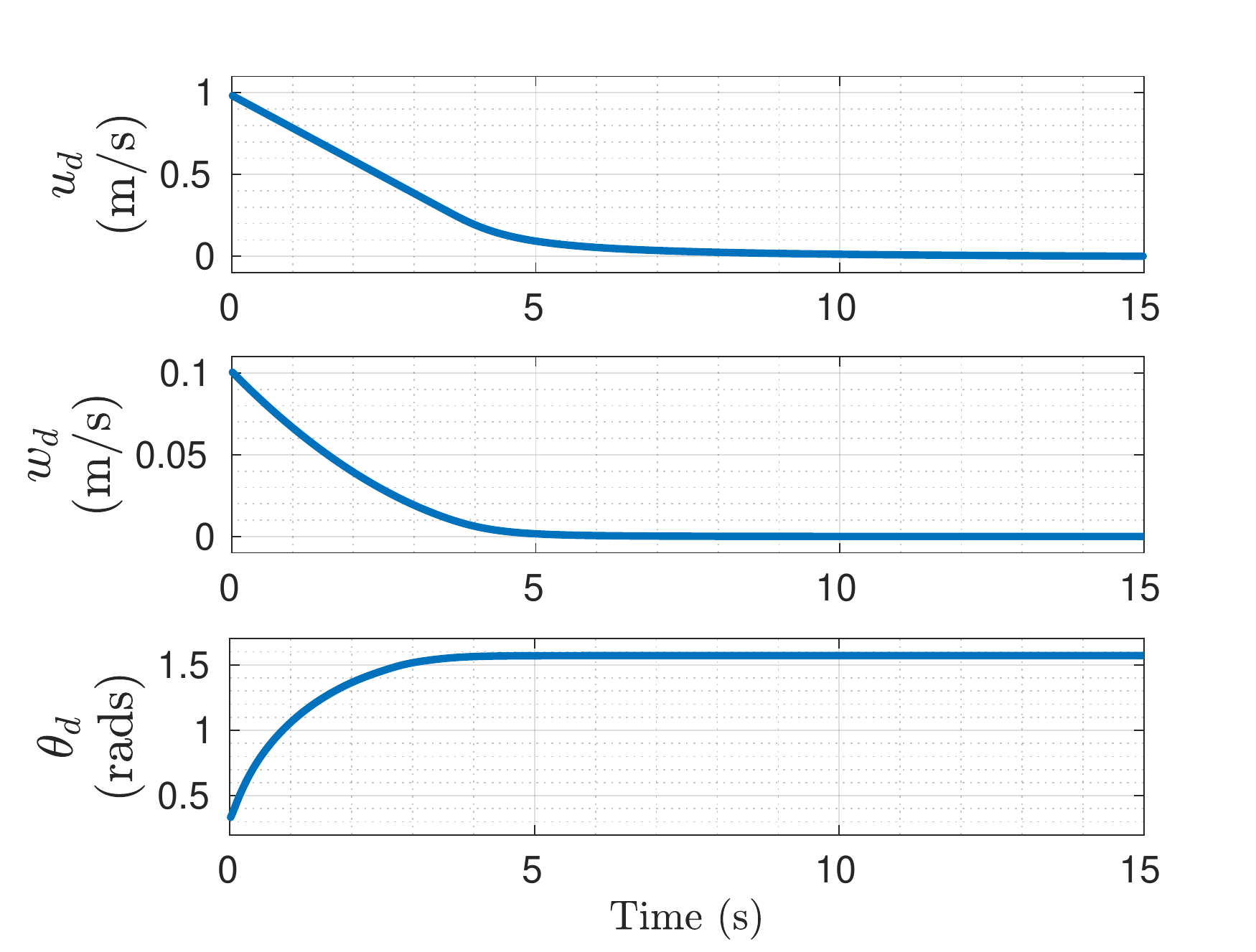}
    \caption{Desired velocities for $u$ and $w$, and desired pitch angle $\theta$ during the transition from cruise to hover. As can be seen, the initial values are relatively high since the UAV is in cruise flight mode and the final speed is almost zero to maintain a static flying.}
    \label{fig:desired_vel_ch}
\end{figure}

\begin{figure}
    \centering
    \includegraphics[width = \columnwidth]{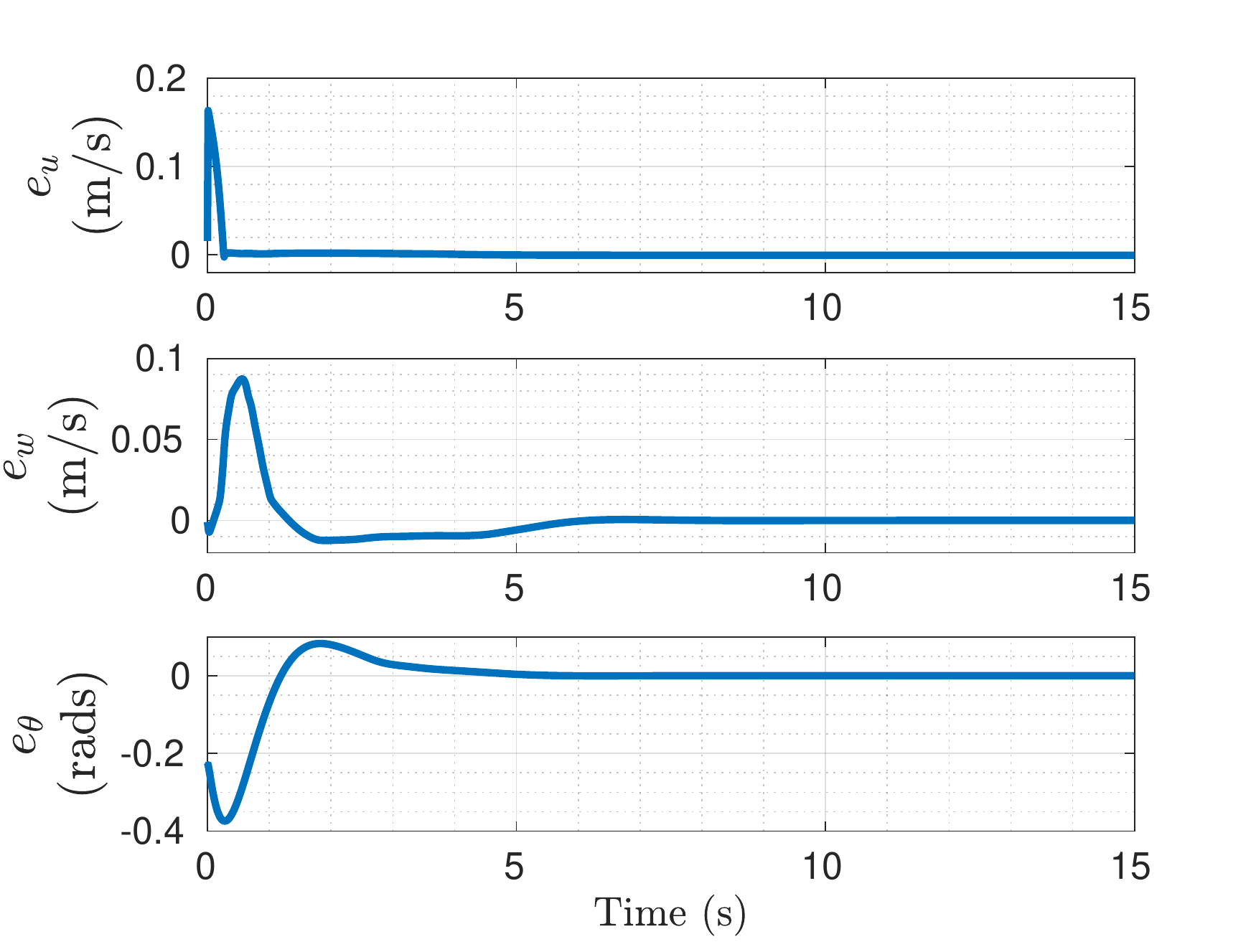}
    \caption{Error velocities $u$ and $w$, and pitch angle error evolution obtained during the simulation for the cruise-hover transition.}
    \label{fig:u_w_ch}
\end{figure}

\begin{figure}
    \centering 
    \includegraphics[width = 0.9\columnwidth]{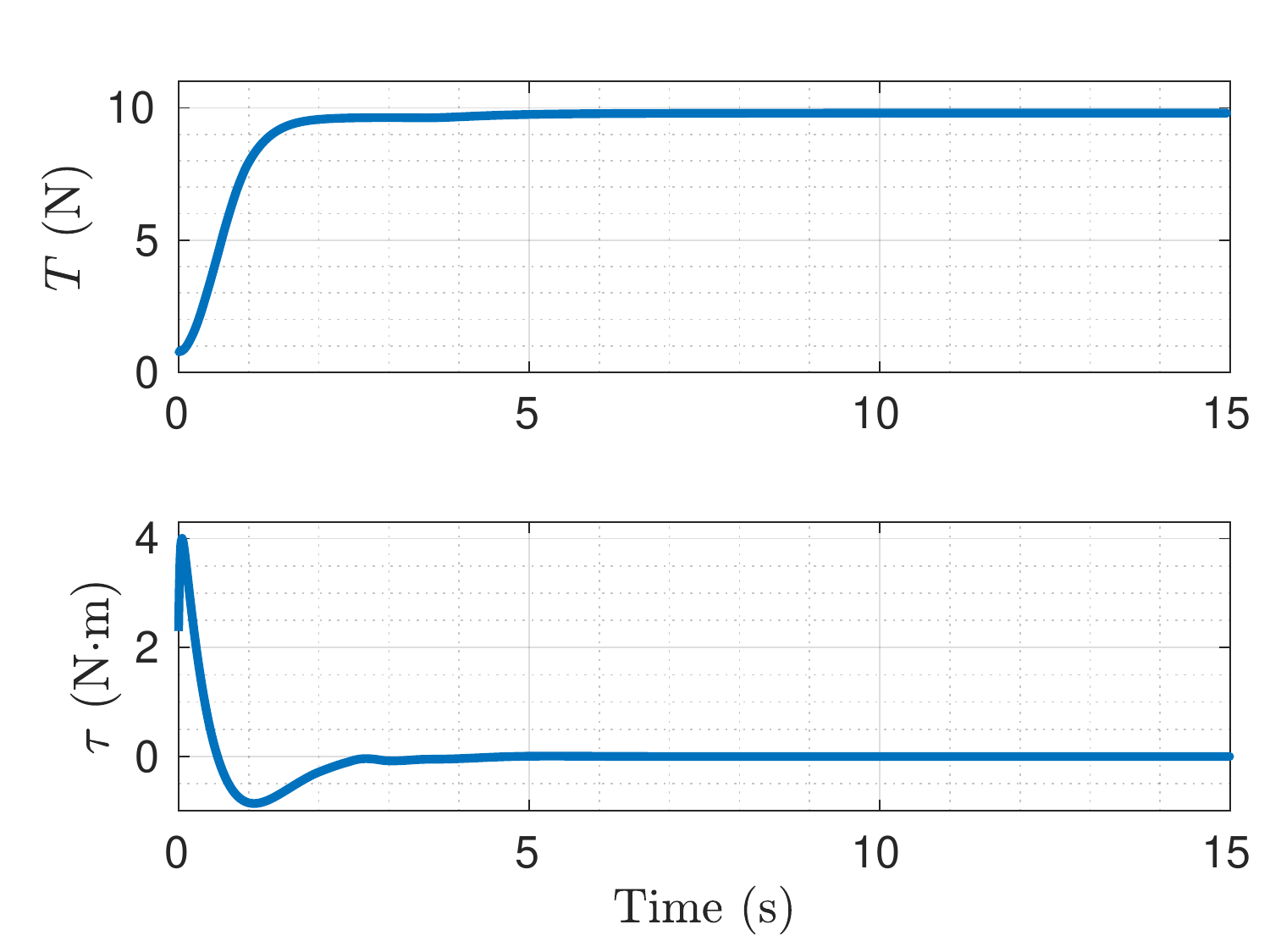}
    \caption{Control input $T$ and $\tau$ applied to the system \eqref{eq:sum1} and \eqref{eq:sum2} for the transition from cruise to cruise}
    \label{fig:thrust_ch}
\end{figure}

For the opposite flight mode transition (cruise-hover), the desired values in the speeds and pitch angles are shown in fig. \ref{fig:desired_vel_ch}. Then the initial conditions of the velocities were $u(0)=1.1$ m/s  and $w(0)=0.16$ m/s and $\theta(0)=0.15$ rads and the results obtained in the implementation of the NN are shown in fig. \ref{fig:u_w_ch} where the velocities also reach the desired value decreasing approximately to zero to keep the hover movement constraints. The control inputs generated during this transition are shown in the fig. \ref{fig:thrust_ch}.

\section{Conclusion}
\label{sec:conclusion}
In this paper, the implementation of an RNN for feedback linearization of nonlinearities in the $u$ dynamics of a tail-sitter is presented, showing through simulations satisfactory results. After linearization of the nonlinearities of the system, a proportional control in thrust for flight mode transition is applied to reach desired velocities and angle during both hover-cruise and cruise-hover transitions. This control approach has some advantages like fast implementation but a principal disadvantage is the computational cost due to its implementation on a micro-controller. As future work we are planning to implement this type of neural network in the estimation of lift forces presented in the use of unknown airfoils and more recently, dynamic airfoils, to verify that the system works with different physical parameters like mass, inertia, etc.



\end{document}